\documentclass{appolb}
\usepackage{graphicx}

\begin{document}
\title{Formation of $\eta'(958)$ bound states by $(\gamma,\it{d})$ reaction
\thanks{Presented at Jagiellonian Symposium on Fundamental
and Applied Subatomic Physics, Krak\'{o}w, Poland, June 7-12, 2015.}%
}
\author{M. Miyatani, H. Nagahiro, S. Hirenzaki
\address{Department of Physics, Nara Women's University, Nara 630-8506, Japan}
\\
\vspace{6.5mm}
{N. Ikeno}
\address{Department of Regional Environment, Tottori University, Tottori 680-8551, Japan}
}
\maketitle

\begin{abstract}
We have investigated the $^6\rm{Li}(\gamma,\it{d})$ reaction theoretically for the formation of the $\eta'(958)$ mesic nucleus. We have reported the numerical results in this article.
\end{abstract}
\PACS{21.85.+d, 36.10.Gv, 25.10.+s, 13.60.Le}
  
\section{Introduction}
The mesic nuclear bound states have been considered to be interesting systems and studied both theoretically and experimentally \cite{Yamazaki2012}. So far, one-nucleon transfer reactions have been mainly considered and successfully used to produce the nuclear bound states of meson which is lighter than nucleon in the recoilless kinematics \cite{Yamazaki2012}. Recently, Ikeno \it et al. \rm studied the two nucleon pick-up reaction on the $^6\rm{Li}$ target for the formation of heavy meson bound states with the $\alpha$ particle \cite{Ikeno_gamma_d}. We develop this study in this article and improve the theoretical method in the following four points;  \\
(i) the distortion effects of the projectile and ejectile are taken into account,  \\
(ii) the elementary cross section is evaluated phenomenologically and the absolute value of the formation cross section is reported,  \\
(iii) the realistic $\alpha$ density distribution is used to calculate the meson-$\alpha$ bound states \cite{Hiyama_private},  \\
(iv) the recoil effects of the formation reaction are considered.  \\
We show some of the numerical results here.

\section{Effective number formalism for the quasi-deuteron in nucleus}
We apply the effective number approach to evaluate the formation rate of the bound systems in the two nucleon pick-up $(\gamma,\it{d})$ reactions as in Ref. \cite{Ikeno_gamma_d}. In the effective number approach, the formation cross section by the $(\gamma,\it{d})$ reaction can be written as,

\begin{eqnarray}
\frac{d^2\sigma}{dE_dd\Omega_d}=\left(\frac{d\sigma}{d\Omega_d}\right)^{\rm{ele}}_{\rm{Lab}}\sum_f N_{\rm{eff}}\frac{\Gamma_{\eta'}}{2\pi}\frac{1}{\Delta E^2+\Gamma_{\eta'}^2/4}, \label{eq:CS}
\end{eqnarray}

\noindent where $\left(d\sigma/d\Omega_d\right)^{\rm{ele}}_{\rm{Lab}}$ is the elementary cross section of the $\eta'$ meson production, and $\Gamma_{\eta'}$ the width of the $\eta'$ bound states. $N_{\rm{eff}}$ is the effective deuteron number. The all combinations of the final states with $^4\rm{He}$ and $\eta'$ are indicated by $\it f$ and are summed up to evaluate the inclusive cross section. The energy transfer of the reaction $\Delta E$ in the laboratory frame is defined as, 

\begin{eqnarray}
\Delta E=T_d-p_\gamma+S_d-B_{\eta'}+m_{\eta'}, \label{eq:delE}
\end{eqnarray}

\noindent where $T_d$ is the emitted deuteron kinetic energy, $p_\gamma$ the incident photon momentum, and $m_{\eta'}$ the $\eta'$ meson mass. The $\eta'$ meson binding energy $B_{\eta'}$ and the deuteron separation energy $S_d$ are determined for each combination of the bound level of $\eta'$ meson and the excited level of the daughter nucleus. Here, we neglect the recoil energy of the nucleus in this expression since we consider the kinematics close to the recoilless condition. The details of the theoretical formula will be reported in Ref. \cite{Miyatani}.

\section{Numerical results}
\begin{figure}[htbp]
\begin{center}
\includegraphics[width=8cm]{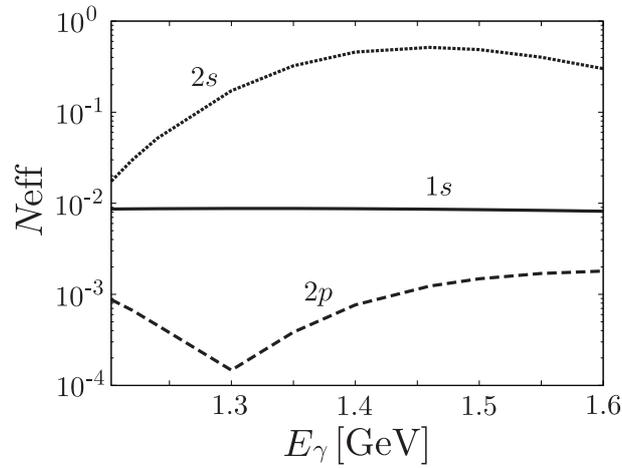}
\end{center}
\vspace{-0.5cm}
\caption{The photon energy dependence of the calculated effective numbers with $(V_0,W_0)=(-150,-5)\,\rm{[MeV]}$. The quantum numbers of the $\eta'$ bound states are indicated in the figure.}
\label{fig:Neff-E1}
\end{figure}

We show in Fig. \ref{fig:Neff-E1} the calculated effective numbers with $(V_0,W_0)=(-150,-5)\,\rm{[MeV]}$ case as an example. For each state, the recoilless condition is satisfied at $E_\gamma\sim1.0\,\rm{[GeV]}$ for the $1s$ state, $E_\gamma\sim1.46\,\rm{[GeV]}$ for the $2s$ state, and $E_\gamma\sim1.30\,\rm{[GeV]}$ for the $2p$ state, respectively. We can see from the figure that the effective number for the $2s$ state formation takes the maximum value at the recoilless energy as expected. On the other hand the effective number for the $2p$ state takes the smallest value at the recoilless kinematics because of the quasi-orthogonal condition between bound $\eta'$ and deuteron wave functions. The effective number for the $1s$ state does not show the clear $E_\gamma$ dependence in this energy region because of the compact wave function of the deeply bound states. Here, the bound $\eta'$ meson wave functions in the final state are calculated by solving the Klein-Gordon equation with the optical potential $U_{\eta'}(r)$ written as,

\begin{eqnarray}
U_{\eta'}(r)=(V_0+iW_0)\frac{\rho_\alpha(r)}{\rho_0},  \label{eq:potential-eta'}
\end{eqnarray}

\noindent where $V_0$ and $W_0$ are the parameters which determine the real and imaginary potential strength, respectively. The normal nuclear density $\rho_0$ is fixed to be $\rho_0=0.17\,\rm{[fm^{-3}]}$.
 
\begin{figure}[htbp]
\begin{center}
\includegraphics[width=12.cm]{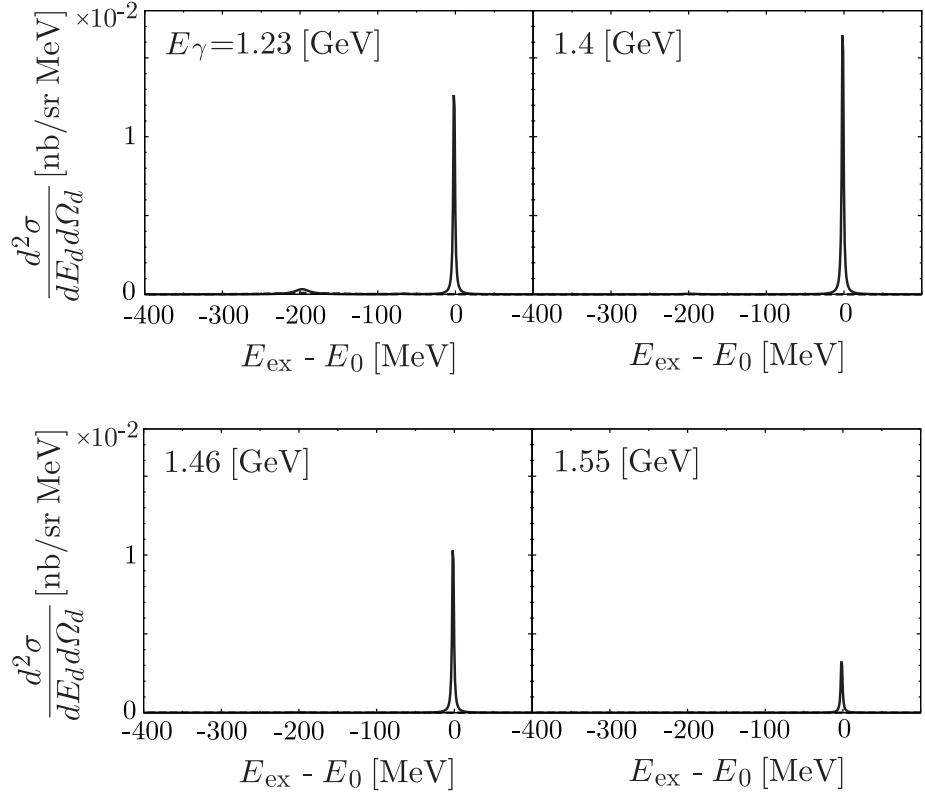}
\end{center}
\caption{Expected spectra of the forward $^6\rm{Li}(\gamma,\it{d})$ reaction for the formation of the $\eta'$ bound state in $\alpha$ are plotted as functions of the excitation energy of the $\eta'$ mesic nucleus for four incident photon energies as indicated in the figure. The parameter of the real part of the optical potential is assumed to be $V_0=-150\,\rm{[MeV]}$ and that of the imaginary part to be $W_0=-5\,\rm{[MeV]}$ for the solid lines. The contribution from the quasi-free $\eta'$ production is not included in these spectra.}
\label{fig:CS-V0_150}
\end{figure}

We show the calculated formation spectra of the $\eta'-\alpha$ bound states in Fig. \ref{fig:CS-V0_150} for the $V_0=-150\,\rm{[MeV]}$ case for four incident photon energies. The potential parameter of the absorptive part is assumed to be $W_0=-5\,\rm{[MeV]}$. The horizontal axis indicates the excitation energy of the $\eta'$ mesic nucleus and the peaks in $E_{\rm{ex}}-E_0<0$ region mean the formation of the bound states.

\clearpage
\section{Conclusions}
We have reported the first calculated results of the formation of the $\eta'(958)$ mesic nucleus in the $^6\rm{Li}(\gamma,\it{d})$ reaction. We show the results for $(V_0,W_0)=(-150,-5)\,\rm{[MeV]}$ case as an example. 

The strength of the $\eta'$-Nucleus potential is still controversial and has not been determined well. In Refs. \cite{Nagahiro2005, Nagahiro2006}, where the formation of the $\eta'$ mesic nucleus was considered for the first time, the strength of the $\eta'$-Nucleus potential was evaluated by NJL model to be around $-150\,\rm{[MeV]}$. Actually the recent evaluation based on the chiral symmetry restoration \cite{Jido2012} also indicates the strong attractive and less absorptive potential. The potential strength considered in this article is based on these studies. Another calculation based on the chiral unitary model \cite{Nagahiro2012} reveals the sensitivity of the potential to the coupling strength of the singlet $\eta$ to the octet baryons. On the other hand, the latest experimental data indicate the small $\eta' N$ scattering length \cite{Czerwinski2014} and the shallow $\eta'$-Nucleus potential \cite{Nanova2013}. Theoretical evaluation in Ref. \cite{Bass2014} also indicates weak attractive potential. Therefore, we believe that the further studies reported in this article in addition to those proposed in Refs. \cite{Nagahiro2013, Itahashi2012} are much important to develop this field further.

The comprehensive results of the $(\gamma,\it{d})$ reaction including those with the shallow $\eta'$-Nucleus potential will be reported in Ref. \cite{Miyatani} soon.

\section*{Acknowledgements}
We acknowledge the fruitful discussions with H. Fujioka and T. Ishikawa from the beginning of this research. We would like to thank E. Hiyama for providing us the realistic density of the $\alpha$ particle. This work is partly supported by the Grants-in-Aid for Scientific Research No. 24540274 (S.H.) and No. 15H06413 (N.I.) in Japan.

\end{document}